\documentclass[%
 reprint,
superscriptaddress,
longbibliography,
 amsmath,amssymb,
 aps,
]{revtex4-2}

\usepackage{graphicx}% Include figure files
\usepackage{dcolumn}% Align table columns on decimal point
\usepackage{bm}% bold math
\usepackage{siunitx}
\usepackage[version=4]{mhchem} % Formula subscripts using \ce{}
\usepackage{placeins}
\usepackage{booktabs} 
\usepackage{tabularx,booktabs}
\usepackage{soul}
\usepackage{hyperref}% add hypertext capabilities
\hypersetup{
    colorlinks,%
    citecolor=blue,%
    linkcolor=blue,%
    urlcolor=blue
}

\usepackage{lipsum}
\usepackage{soul}            % for strikethrough \st{}
\usepackage{wrapfig}
\usepackage[dvipsnames]{xcolor}  % color definitions for editing macros (see below)
\usepackage{hyperref}
\hypersetup{colorlinks, citecolor=blue, linkcolor=blue, urlcolor=blue}
\usepackage[caption=false]{subfig}
\usepackage{subfig}
\usepackage{ulem}

\newcommand{\orcid}[1]{\href{https://orcid.org/#1}{\includegraphics[width=8pt]{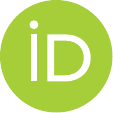}}}

\DeclareSIUnit\rydberg{\text {Ry}}

\begin{document}

\preprint{APS/123-QED}

\title{Exploring Topological Transport in Pt$_2$HgSe$_3$ Nanoribbons: Insights for Spintronic Device Integration}

\author{Rafael L. H. Freire\orcid{0000-0002-4738-3120}}
\email{rafael.freire@lnnano.cnpem.br}
\affiliation{Ilum School of Science, Brazilian Center for Research in Energy and Materials (CNPEM), Campinas, SP, Brazil.}
 
\author{F. Crasto de Lima\orcid{0000-0002-2937-2620}}%
\email{felipe.lima@ilum.cnpem.br}
\affiliation{Ilum School of Science, Brazilian Center for Research in Energy and Materials (CNPEM), Campinas, SP, Brazil.}

\author{Roberto H. Miwa\orcid{0000-0002-1237-1525}}
\email{hiroki@ufu.br}
\affiliation{ Instituto de Física, Universidade Federal de Uberlândia, 38400-902 Uberlândia, Minas Gerais, Brazil.}

\author{A. Fazzio\orcid{0000-0001-5384-7676}}
\email{adalberto.fazzio@ilum.cnpem.br}
\affiliation{Ilum School of Science, Brazilian Center for Research in Energy and Materials (CNPEM), Campinas, SP, Brazil.}

\date{\today}

\begin{abstract}

The discovery of the quantum spin Hall effect led to the exploration of the electronic transport for spintronic devices. Here, we theoretically investigated the electronic conductance in large-gap realistic quantum spin Hall system, Pt$_2$HgSe$_3$ nanoribbons. By an ab initio approach, we found that the edge states present a penetration depth of about $0.9$\,{nm}, which is much smaller than those predicted in other 2D topological systems. Thus, suggesting that Pt$_2$HgSe$_3$ allows the exploitation of topological transport properties in narrow ribbons. Using non-equilibrium Green's functions calculations, we have examined the electron conductivity  upon the presence of Se\,$\leftrightarrow$\,Hg antistructure defects randomly distributed in the Pt$_2$HgSe$_3$ scattering region. By considering scattering lengths up to $109$\,nm, we found localization lengths that can surpass $\mu$m sizes for narrow nanoribbons ($<9$\,nm). These findings can contribute to further understanding the behavior of topological insulators under realistic conditions and their integration within electronic, spintronic devices.

\end{abstract}

\maketitle

\section{\label{sec:level1} Introduction}

Topological insulators (TIs) are an emerging class of intriguing materials with unique electronic properties. Particularly, the wave function that describes their electronic states spans a Hilbert space with a new topology. The consequence is that at any interface, with an ordinary insulator, they will present a gapless state protected by time-reversal symmetry  \cite{RMPhasan2010}. 

The first indirect observation of the topological edge states was the measurement of the quantized conductance in HgTe/CdTe quantum wells \cite{SCIENCEkonig2007, SCIENCEroth2009}. On the other hand, the direct observation of the electronic state on the surface was possible through an angle-resolved photoemission spectroscopy (ARPES) experiment, first in \ce{Bi_{1-x}Sb_x} \cite{NATUREhsieh2008}. Despite this, characterizing the surface structure of such materials is not an easy task, and later combined measurements of ARPES, scanning tunneling microscopy (STM), and scanning tunneling spectroscopy (STS) helped to identify TI materials \cite{PRLalpichshev2010, BOOKwang2016}.

Measurements of quantized conductance in the quantum spin Hall geometry have been done in quantum well systems HgTe/CdTe and InAs/GaSb \cite{SCIENCEkonig2007, SCIENCEroth2009, PRLrui2015}, on decorated/interfaced graphene \cite{SCIADVhatsuda2018}, and on WTe$_2$ \cite{SCIENCEwu2018}.
STM/STS of edge states were also observed on WTe$_2$ \cite{NATPHYStang2017, NATCOMMchen2018}, SiC/Bismuthene interface \cite{SCIENCEreis2017} and in jacutingaite mineral (Pt$_2$HgSe$_3$) \cite{NLkandrai2020}. However, those systems are either quantum wells, low gap decorated graphene, and/or 2D materials unstable under oxidation \cite{SMALLye2016, 2DMATnaylor2017, PCCPfreire2024}. In other words, only a few 2D materials with the topological helical edge mode showing quantized transport have been measured. This drives a quest to deeply understand their structural and energetic stability, how intrinsic defects, confinement length (ribbon width), and temperature, to cite a few, can affect the quantized transport.

Jacutingaite (Pt$_2$HgSe$_3$) is a naturally occurring mineral \cite{TNcabral2008}, stable against oxidation \cite{NLkandrai2020}. It gained attention due to the large spin-orbit coupling effect leading to a non-trivial $0.15$\,eV energy gap opening \cite{PRLmarrazzo2018, PRBlima2020}, being the first realization of the Kane-Mele topological model \cite{PRLkane2005}. While it is a chemically stable topological insulator, the experimental synthesis of Jacutingaite is prone to present defects and disorders. Such effects in the electronic transport of the topological states, combined with confinement potential, bulk density of state, and topological state penetration depth, can change the quantized topological conductance.

In this work, we systematically study the ballistic transport in jacutingaite nanoribbons including stoichiometric defects, by combining density functional theory (DFT) and electronic transport calculations through non-equilibrium Green's functions (NEGFs). We explored different concentrations of non-magnetic defects in a disordered geometry and the competition between topological edge-state manifestation and confinement in narrow ribbons. Here we showed that the localization length in the transport can surpass $\mu m$ length, that is, allowing realistic device engineering in the jacutingaite platform despite the presence of intrinsic defects.

\section{Methodology}

\subsection{Equilibrium geometry and electronic structure}

To obtain the geometric, energetic, and electronic properties of pristine and defective systems, we performed density functional theory (DFT) calculations as implemented in the plane wave basis Vienna Ab-Initio Simulation Package (VASP) \cite{PRBkresse1993, PRBkresse1996}. We employed the semilocal exchange-correlation formalism proposed by Perdew, Burke, and Ernzerhof (PBE) \cite{PRLperdew1996, PRLperdew1997}. To improve the description of van der Waals interactions, we used the DFT-D3 pairwise corrections proposed by Grimme  \cite{JCPgrimme2010}. The plane-wave basis energy cutoff was set to $400$\,eV for all calculations with an energy convergence parameter of $10^{-6}$\,eV, and atoms were allowed to relax until all forces were smaller than $10^{-3}$\,eV/{\AA}. The electron-ion core interactions were described through the projected augmented wave (PAW) potentials \cite{PRBblochl1994, PRBkresse1999}. We integrated the Brillouin zone through a $4\times4\times1$ k-point mesh for relaxation. We further included relativistic effects by considering spin-orbit coupling corrections (SOC).

\subsection{Electronic transport}

%%%%%%FIG1
\begin{figure}
\includegraphics[width=0.45\textwidth]{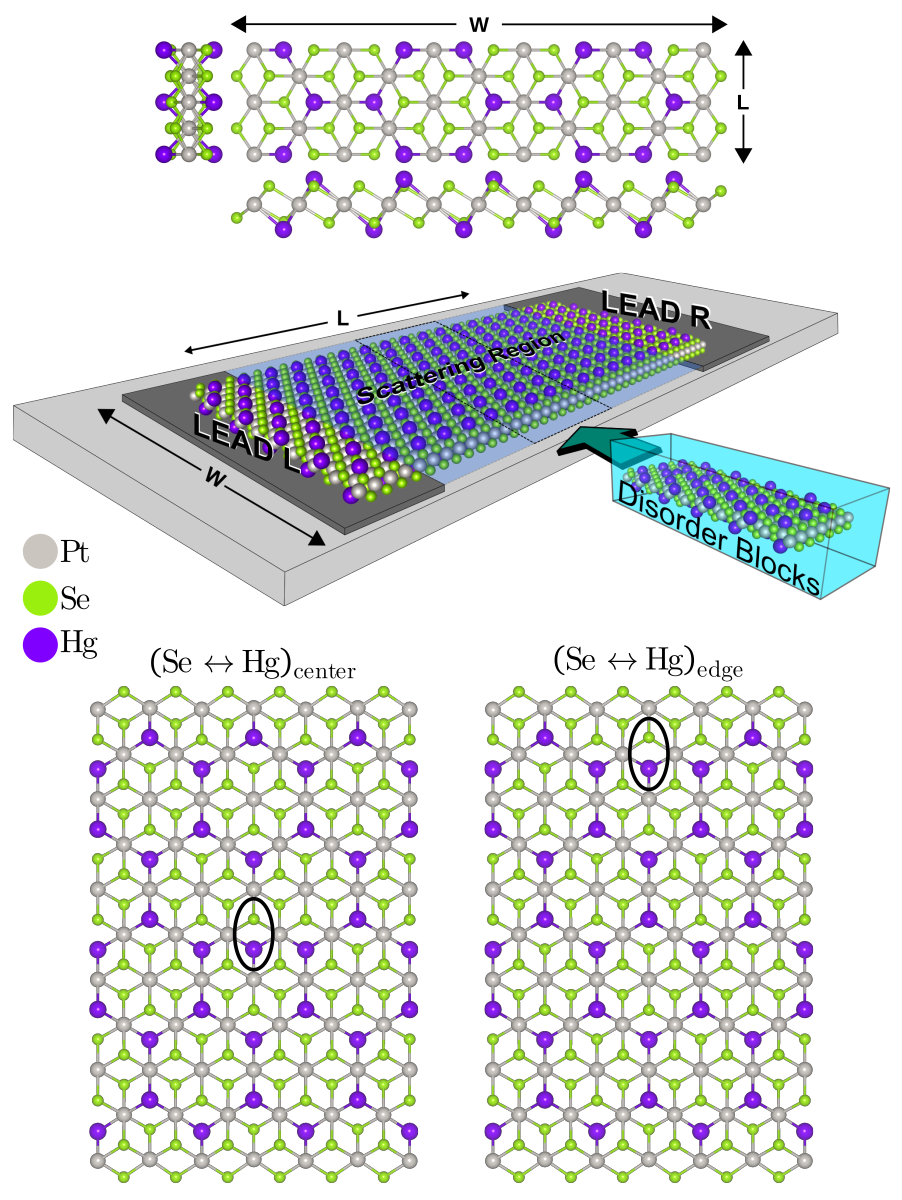}
\caption{\label{fig:nrdim} (up) Nanoribbon dimensions and illustrative device for electronic transport measurement. (bottom) Defective building blocks indicating center and edge antistructure defects. Gray, (purple/black), and (green/red) atoms represent \ce{Pt}, \ce{Hg}, and \ce{Se} atoms, respectively.}
\end{figure}

In Fig.\,\ref{fig:nrdim}, we present the setup used for the electronic transport calculation. We have considered the Se-terminated zigzag nanoribbons (NRs)\,\cite{NLkandrai2020} where the electronic transport takes place along the $z$-direction, and three NR widths ($W$) of $34$, $60$, and $99$\,{\AA}, parallel to the $x$-direction. We have combined DFT and a recursive Green's function (RGF) method \cite{PRLrocha2008}. The system can be decomposed into three main parts, two semi-infinite charge reservoirs, that are the ($i$) left and ($ii$) right electrodes, and ($iii$) the central scattering region. Besides, within the RGF the scattering region can be further split into several building blocks (Disorder Blocks in Fig.\,\ref{fig:nrdim}). Two distinct building block kinds have been examined, as shown in Fig.\,\ref{fig:nrdim}(bottom), one with the anti-structure defect located in the NR's central region and the other with the defect close to the NR's edge sites. The concatenation
of these blocks along the transport-direction is random, resulting in scattering lengths ($L$) up to $0.1$\,{$\mu$m}. It is also important to note that the blocks were constructed to ensure coupling only between adjacent neighbors. For the transport calculations, we first obtain the electrodes and scattering region Hamiltonians through SIESTA-based DFT calculations. Thus, we employed norm-conserving Troullier-Martins pseudopotentials \cite{PRBtroullier1991}, and single $\zeta$-polarized (SZP) basis set of numerical atomic orbitals to expand the Kohn-Sham orbitals of the valence electrons. Our real-space mesh cutoff was set to $300$\,{Ry} and the Brillouin zone was integrated through a Monkhorst-Pack $k$-points mesh $1 \times 1 \times 10 $ ($1\times 1 \times 3 $) for the electrodes (scattering) region. Our SZP basis set calculations agree with our plane-wave-based calculations, thus being sufficient to describe the electronic structure of the system.

\section{Results}

Although the present study focuses on the topologically protected electronic transport in Pt$_2$HgSe$_3$ nanoribbons, we will start our investigation by examining key structural and electronic properties of pristine and defective Pt$_2$HgSe$_3$ monolayer.

\subsection{Pt$_2$HgSe$_3$ monolayer}

\subsubsection{Pristine system}

%%%%%%FIG2
\begin{figure}
\includegraphics[width=0.9\columnwidth]{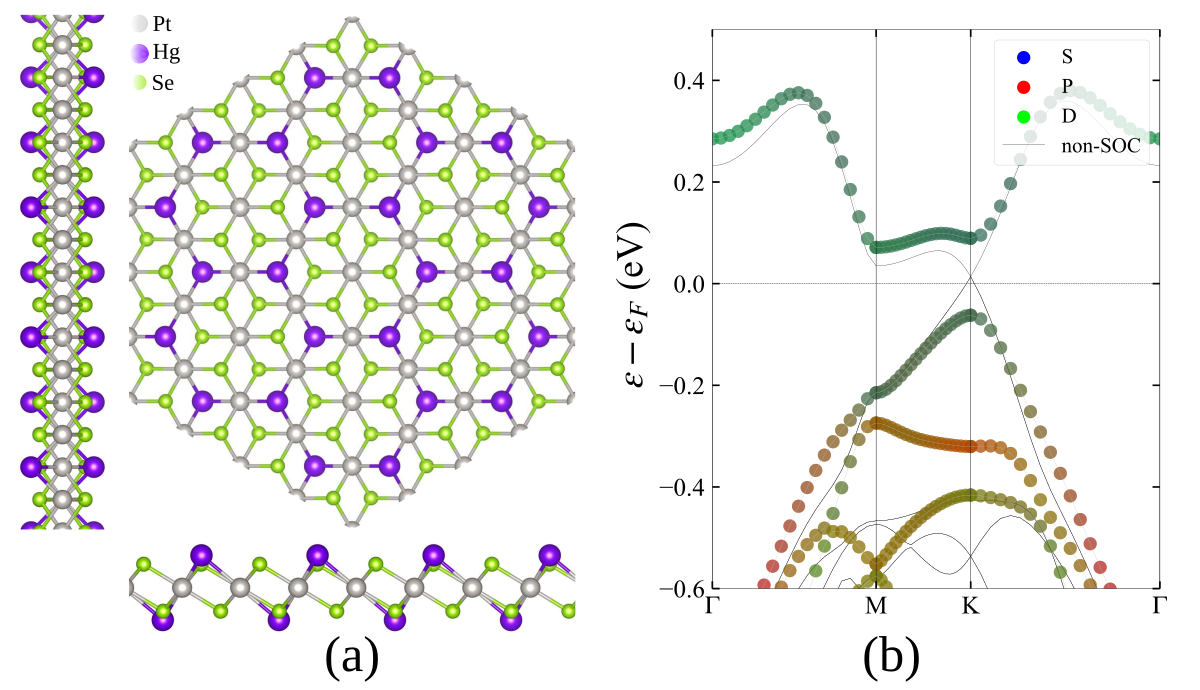}
\caption{\label{fig:MLbandstr} (a) Jacutingaite monolayer top and side views and (b) electronic band structure with and without SOC effects. Gray, purple, and green atoms represent Pt, Hg, and Se atoms, respectively.}
\end{figure}

The Hg atoms in Pt$_2$HgSe$_3$ pinpoint a honeycomb structure, purple atoms in Fig.~\ref{fig:MLbandstr}(a), embedded in a PtSe matrix. At the equilibrium geometry, the electronic band structure obtained without the inclusion of the spin-orbit coupling (SOC) is characterized by a Dirac cone structure crossing the Fermi level at the K-point, Fig.~\ref{fig:MLbandstr}(b). Meanwhile, upon the inclusion of the SOC, we found an energy gap at the Dirac point with the band edges composed mainly by  Hg-$6s$ orbitals hybridized with Pt-$4d$ orbitals.  Such behavior is typical of the Kane-Mele model \cite{PRLkane2005}, and characterizes the topological phase of jacutingaite \cite{PRLmarrazzo2018, PRBlima2020}. The energy gap opened by the spin-orbit coupling in the pristine system is calculated here as $0.13$\,eV. It is worth pointing out that previous studies considering hybrid functionals (HSE) and GW calculations predicted a higher energy gap of $0.2$\,eV \cite{PRBlima2020, PRBwu2019} and $0.5$\,eV \cite{PRLmarrazzo2018}. However, experimental STM measurements at $9$\,K estimate the energy gap to be $0.08 \pm 0.03$\,eV \cite{NLkandrai2020}, closer to the one obtained by using the DFT-GGA approach. 

\begin{table}
\begin{ruledtabular}
\caption{\label{table:defects} Intrinsic point defects' formation energies.}
    \begin{tabular}{cc}
    Defect           & $E^f$ (\si{\eV})     \\
    \hline
    V$_{\rm Se}$        & $5.42$ \\
    V$_{\rm Pt}$        & $6.87$ \\
    V$_{\rm Hg}$        & $0.23$ \\
    $({\rm Se \leftrightarrow Hg})$ & $0.79$ \\
    \end{tabular}
\end{ruledtabular}
\end{table}

\subsubsection{Intrinsic defects}

The emergence or suppression of a given topological phase as well as its influence on electronic transport can be dictated by the presence of defects\,\cite{NATCOMMlpke2017, PRBfukui2020, NLlima2021, SSCpezo2023}. For instance, the trivial\,$\rightarrow$\,non-trivial QSH transition is mediated by the concentration of selenium vacancies (V$_{\rm Se}$) in  PtSe$_2$ ML\,\cite{NLlima2021}.

Here, we have considered vacancies (V$_{\rm X}$, with X = Se, Pt, and Hg), and antistructure (Se\,$\leftrightarrow$\,Hg) defects \cite{PRBjanotti1997} in Pt$_2$HgSe$_3$\,ML. The occurrence rate of these defects can be inferred by the calculation of the formation energy ($E^f$)\cite{komsa2012two,chen2022atomic},
\begin{equation}
  E^f = E_{\rm pristine} - E_{ \rm defect} - n\times E_{\rm X},  
\end{equation}
where $E_{\rm pristine}$ and $E_{\rm defect}$ are the total energies of the pristine and the defective Pt$_2$HgSe$_3$ ML; $E_{\rm X}$ is the total energy of an isolated atom X, and $n$ is the number of missing atoms.  For a single X vacancy, V$_{\rm X}$,  $n$\,=\,1, while for the antistructure defect, which is stoichiometric, $n$\,=\,0. In the antistructure defects, we have considered antisites, Se$_{\rm Hg}$ and Hg$_{\rm Se}$,  created adjacent to each other.  Generally, transition metals vacancy have larger defect formation energy than the chalcogen vacancy \cite{PRMfreire2022}, which is also verified here, as V$_{\rm Pt}$ formation energy is higher than V$_{\rm Se}$. However, the formation energy of both defects is larger compared with V$_{\rm Hg}$, $0.23$\,eV, followed by the (Se\,$\leftrightarrow$\,Hg) antistructure, 0.79\,eV.

The stoichiometry of Pt$_2$HgSe$_3$ has been conserved in experimentally synthesized jacutingaite samples \cite{TCMvymazalova2012, NLkandrai2020}. Meanwhile, naturally occurring samples exhibit Pd doping in the Pt site while maintaining the Hg/Se ratio\,\cite{TNcabral2008}. These results, combined with the lower formation of the  (stoichimetric) antistructure defects, allow us to infer that (Se\,$\leftrightarrow$\,Hg) interchanged defect is quite likely to be present in Pt$_2$HgSe$_3$\,\cite{2DMlima2023}. In the sequence, we will examine the effect of the local disorder induced by the antistructure defect on the electronic transport properties, mediated by the topologically protected edge states in Jacutingaite NRs.

\subsection{Pt$_2$HgSe$_3$ nanoribbons: electronic transport}

The transport geometry for the monolayer Pt$_2$HgSe$_3$ is presented in Fig.~\ref{fig:nrdim}. The leads are pristine systems, and the scattering region contains the antistructure defects. We explored ($i$) different nanoribbons width ($W=34$, $60$, and $99$\,{\AA}), ($ii$) different antistructure concentrations ($\lambda = 1.9 \times 10^6$, and $4.5 \times 10^6$\,defects/cm), and ($iii$) different scattering region length ($L=15.6$, $31.2$ and $109.2$\,nm).

\subsubsection{Pristine nanoribbons}

%%%%%%FIG3
\begin{figure}
\includegraphics[width=0.4\textwidth]{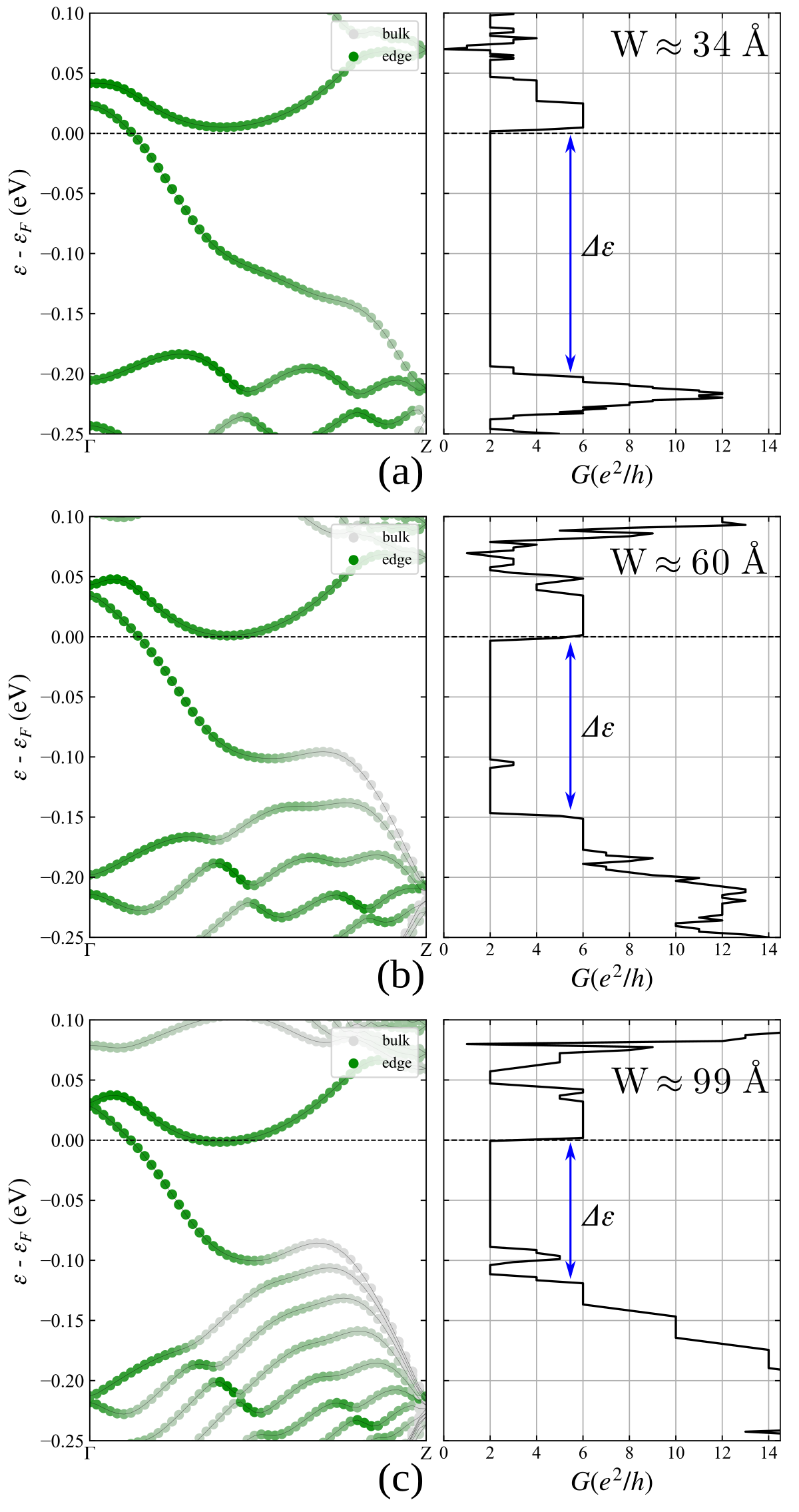}
\caption{\label{fig:trcprist} Electronic band structure projected on the bulk and edge regions (left) and conductance (right) across a Pt$_2$HgSe$_3$ nanoribbon with different size width, (a) $W=34$, (b) $W=60$ and (c) $W=99$\,{\AA}. The green and gray colors in the band structure is proportional to the edge ($|\langle {\rm edge} | n,\vec{k}  \rangle|^2$) and bulk ($|\langle {\rm bulk} | n,\vec{k}  \rangle|^2$) contribution to the system eigenstate, respectively. For all band structures, each NR edge region for projection is about $7.69$\,{\AA} wide (perpendicular to transport direction), while the remaining atoms are considered as bulk projection region.}
\end{figure}

First, we analyze the conductance through a defect-free jacutingaite NR. In such a case, the incident wave function is also an eigenstate of the scattering region, thus leading to a suppression of scattering processes. In Figs.\,\ref{fig:trcprist}(a)-(c), we present the electronic band structure and the conductance  results (in units of $G_0$\,=\,$e^2/h$) for different NR widths, $W=34$, $60$, and $99$\,{\AA}. The band structure presents the projected contribution of the atomic orbitals from the middle of the ribbon, namely bulk-like region $|\langle {\rm bulk} | n,\,\vec{k} \rangle|^2$, and the orbitals at the edge $|\langle {\rm edge} | n,\,\vec{k} \rangle|^2$. We can identify two topological edge bands close to the Fermi energy. For $W=34$, and $60$\,{\AA}, we can see a gap opening at the Dirac crossing due to the inter-edge interaction. In contrast, for $W=90$\,{\AA},  we find the emergence of a Dirac cone at the $\Gamma$-point.

As shown in Fig.\,\ref{fig:trcprist}(right), the energy window ($\Delta\varepsilon$) close to the Fermi level with $G$\,=\,$2G_0$, due to the edge metallic band,  decreases as the NR width increases. That is, for $W=90$\,{\AA} [Fig.~\ref{fig:trcprist}(c)] we found $\Delta\varepsilon$ of around $0.13$\,eV, which is compatible with the topological band gap of the single layer jacutingaite. Whereas, for $W=34$ and $60$\,{\AA} [Figs.\,\ref{fig:trcprist}(a) and (b)], in addition to the energy gap at the Dirac point due to the inter-edge interactions, $\Delta\varepsilon$ increases to about $0.19$ and $0.15$\,{eV}, which can be attributed to the confinement potential sparsing the bulk-like bands \cite{PRLson2006}.

Since the  inter-edge interactions are mediated by the bulk (inner) states, the penetration length ($\ell$) of the edge states is a key quantity in understanding the topologically protected electronic transport along the NR's edge channels. Within the Kane-Mele model, the penetration depth is $\ell\sim at\sqrt{3}/(2\lambda)$, with $a$ the lattice parameter, $t$ the nearest-neighbor hopping, and $\lambda$ the spin-orbit strength term. Taking these values from the jacutingaite band structure with $t/6$ the Dirac band width at $\Gamma$-point, the SOC gap at K-point, $E^\text{K}_\text{gap} = 6\sqrt{3}\lambda$, and $a=7.5$\,\AA, we can estimate $\ell=11.2$\,nm. In contrast, in Ref.\,\cite{PRLmarrazzo2018} the authors predicted a penetration depth ($\ell=\hbar v_F/E_\text{gap}$) of about $4.7$\,\AA, by using an energy gap $E_\text{gap}$ of $0.53$\,eV, larger than that observed experimentally, and $v_F$ of $3.6 \times 10^{5}$\,{m/s}. Further STM measurements, indeed, observed a penetration length of $\sim$\,5\,\AA\, in the Se terminated zigzag jacutingaite NRs\,\cite{NLkandrai2020}.

%%%%%%FIG4
\begin{figure*}
\includegraphics[width=0.95\textwidth]{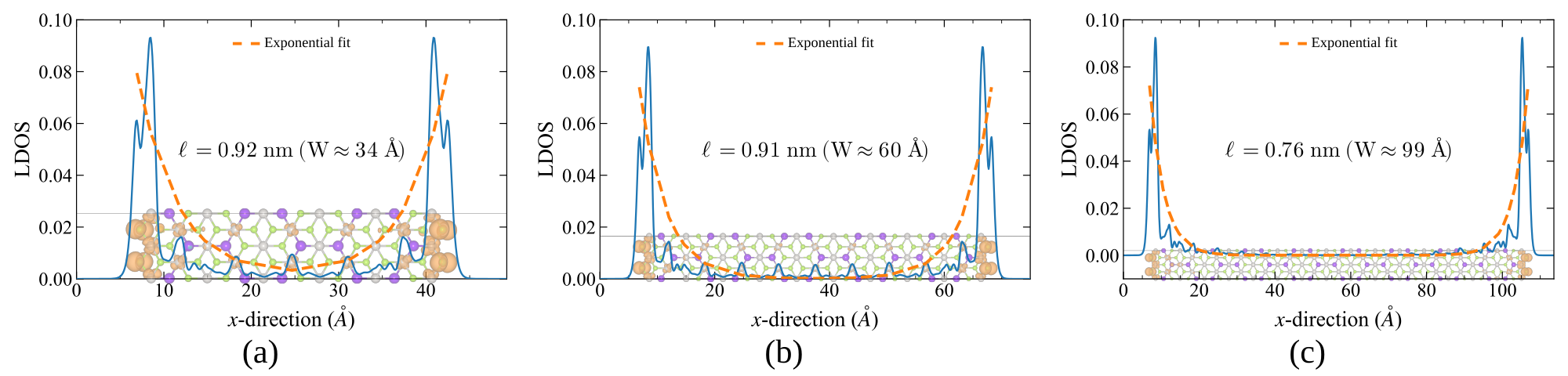}
\caption{\label{fig:ptlength} Penetration depth estimation for different NR-width. Blue curves indicate the $yz$-planar averaged local density of states (LDOS) according to their real-space projection as depicted for each NR-width, obeying $|\Psi_{\rm top} (x)|^2 = |\phi(x)|^2 e^{-2x/\ell}$ . Orange dashed lines are the exponential fitting to obtain the respective penetration depth $\ell$.}
\end{figure*}

By projecting the edge states contribution in real space, Fig.~\ref{fig:ptlength}, we can explicitly calculate the penetration depth ($\ell$) by fitting the wave function decay as $\Psi_{\rm top}(x) = \phi(x) e^{-x/\ell} $, where $\phi(x)$  is the planar ($yz$) averaged single-particle wave-function (near the $\Gamma$-point) of the metallic band crossing the Fermi level.  We found a mean value $\ell=0.91$\,nm for $W=34$ and $60$\,\AA\, [Figs.\,\ref{fig:ptlength}(a) and (b)], and $\ell=0.76$\,nm for the widest NR, $W=99$\,\AA. It is worth noting that these values of $\ell$ are lower than predicted considering the empty honeycomb lattice (Kane-Mele model), $11.2$\,nm, suggesting that the PtSe$_2$ background's dielectric media improves the wave function's screening, and thus reducing penetration depth. 

The reduced penetration depth implies that the topological electronic transport is maintained against inter-edge scattering processes even in narrow jacutingaite NRs.  However, in the case of scattering potentials along the NRs \cite{konig2013spatially, essert2015magnetotransport}, we might observe an entirely different picture. In this scenario, the inter-edge interactions may lead to the suppression of the electronic transport along the edge channels.

\subsubsection{Antistructure defects in NRs}

\begin{figure}
\includegraphics[width=0.9\columnwidth]{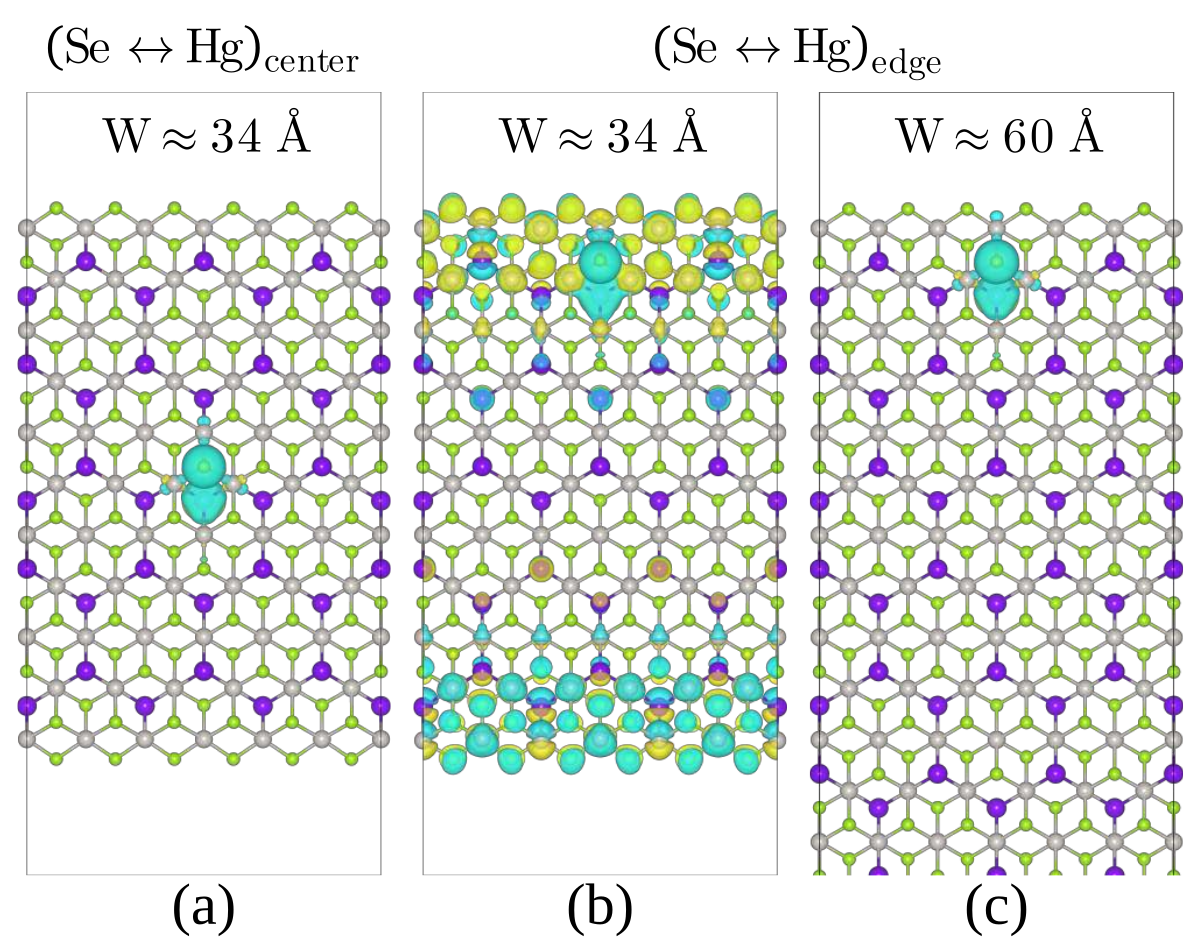}
\caption{\label{fig:diffcharge} Charge density difference with/without the defect presence $\Delta \rho (\vec{r}) = \rho_{\rm def}(\vec{r}) - \rho_{\rm pris}(\vec{r})$, for the $W=34$\,{\AA} nanoribbon (a) and (b) and for $W=60$\,{\AA} nanorribon in (c). In (a) the defect is located in the center of the ribbon, while in (b) and (c) the defect is at the ribbon edge. Blue/yellow regions indicate a decrease/increase in the electron concentration. The isosurface value is 0.01\,$e$\,\AA$^{-1}$}
\end{figure}

Here, we will study the effect of the scattering potentials induced by the (Se\,$\leftrightarrow$\,Hg) antistructure on the electronic transport properties along the jacutingaite NRs. In Figs.~\ref{fig:diffcharge}(a) and (b), we present the total charge rearrangement ($\Delta\rho$), defined as the total charge difference between the defective and pristine (defect-free) systems,
$\Delta\rho = \rho_\text{defect} - \rho_\text{prist}$, induced by the antistructure defect at the center, and at the edge sites of the NR with $W=34$\,\AA. In the former,  we see a charge rearrangement in diameter of $\sim 0.7$\,nm, giving a scale of the antistructure defect localization. In contrast, as depicted in Fig.\,\ref{fig:diffcharge}(b), an antistructure defect at the edge site of the NR leads to a charge rearrangement in both edges, ruled by the edge-edge coupling present for such width. Meanwhile, by increasing the ribbon width to $W=60$\,{\AA} [Fig.~\ref{fig:diffcharge}(c)], although both NRs present practically the same penetration length, $\ell=0.9$\,nm [Figs.\,\ref{fig:ptlength}(a) and (b)], such a charge density rearrangement at the opposite side to the defect becomes nearly absent. That is, as expected, the inter-edge interactions are weakened for larger values of NR width with respect to the penetration length.

%%%%%%FIG5

\begin{figure*}[t]
\includegraphics[width=0.9\textwidth]{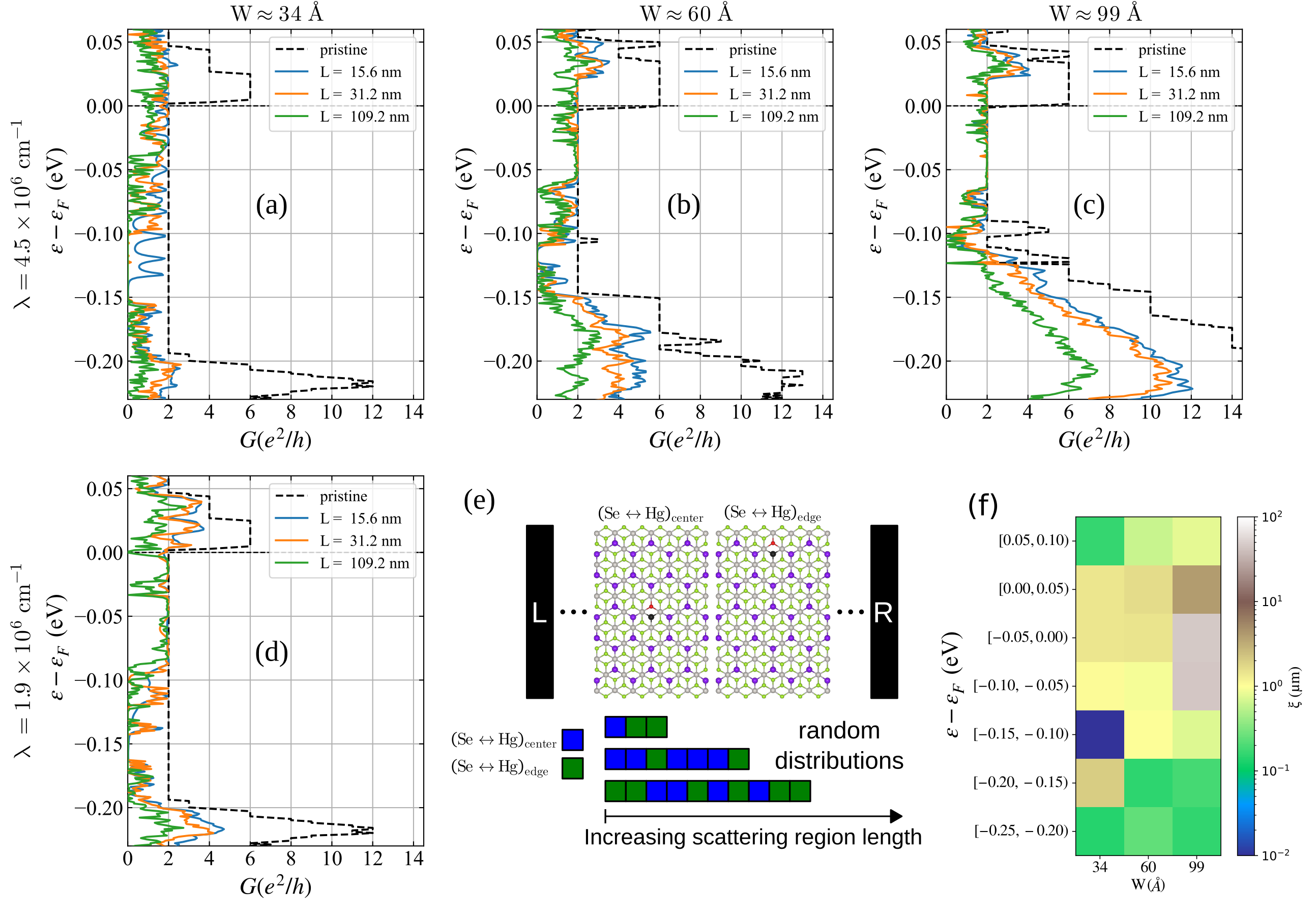}
\caption{\label{fig:conduct} Conductance as a function of nanoribbon length for defect scattering region. (a)-(c) for linear defect concentration of $\lambda = 4.5 \times 10^6$\,cm$^{-1}$ and (d) for $\lambda = 1.9 \times 10^6$\,cm$^{-1}$. (a) and (d) the ribbons have width $W=34$\,{\AA} , (b) $W=60$\,{\AA} and (c) $W=99$\,{\AA}. In panel (e) we schematically present the distribution of the defects as a function of the ribbon length. (f) The average localization lenght $\xi$ for different energy ranges and nanorribon with $W$ for $\lambda = 4.5 \times 10^6$\,cm$^{-1}$.}
\end{figure*}

Focusing on the electronic transport properties, using  the simulation setup presented in Fig.~\ref{fig:nrdim}, we calculate the conductance, $G$, as a function of the width, $W$, and length ($L$) of the NR. We have considered scattering lengths ranging from $16$ up to $109$\,nm with the antistructure defects randomly distributed along the scattering region, as described in Section II-B, and schematically depicted in Fig.~\ref{fig:conduct}(e).

In Figs.~\ref{fig:conduct}(a)-(c), we present our results of conductance at the presence of antistructure defects ($G_\text{defect}$, in units of $G_0=e^2/h$) near the Fermi level,   for $W=34$, $60$, and $99$\,{\AA} and linear concentration of defects ($\lambda$) of $4.5\times 10^{6}$\,{cm$^{-1}$}. It is noticeable that $G_\text{defect}$  is smaller than the conductance of the pristine NR ($G_\text{prist}$) indicated by dashed lines in Fig.~\ref{fig:conduct}. Since the antistructure defects are non-magnetic, the backscattering process in a single edge is forbidden by the time-reversal symmetry, thus we can infer that the results of $G_\text{defect} < G_\text{prist}$ are due to the partial or total inter-edge backscattering processes within the energy window ($\Delta\varepsilon$ in Fig.~\ref{fig:trcprist}) of the topological edge states. It is observed that the reduction of $G_\text{defect}$ with respect to $G_\text{prist}$ becomes more pronounced, independent of the NR width, for energies below $-0.1$\,eV, i.e. $\varepsilon-\varepsilon_\text{F}<-0.1\,\text{eV}$, suggesting the predominance of the bulk states in the electronic transport along the NR. Furthermore, the following results are noteworthy:

($i$) For a given NR width and defect concentration, $\lambda$, $G_\text{defect}$ decreases for higher values of $L=15.6 \rightarrow 109.2$\,nm, since the scattering rate increases with the length of the scattering region. Indeed, as shown in Figs.\,\ref{fig:conduct}(a) and (d), by reducing the concentration of the defects, $\lambda=4.5\rightarrow 1.9\,\times 10^6$\,{cm$^{-1}$}, we find that the reduction of $G_\text{defect}$ is mitigated.

($ii$) For a given value of $\lambda$, the reduction of \,$G_\text{defect}$ is also mitigated by increasing the NR's width, $W=34 \rightarrow 99$\,{\AA} [Figs.\,\ref{fig:conduct}(a)\,$\rightarrow$\,(c)]. This can be attributed to the reduction of the inter-edge backscattering since the ratio between the penetration length of the edge states and the width of the NR, $\ell/W$, reduces for wider systems. For instance, $\ell/W = 0.27$ and $0.08$ for $W=34$ and $99$\,{\AA}, respectively.

The energy dependence of $G_\text{defect}$ and the width of the NRs [($ii$)] reveals that the deviation of $G_\text{defect}$ with respect to $G_\text{prist}$ is proportional to the inter-edge orbital couplings mediated by the antistructure states. In the present study of antistructure defects, we find that $G_\text{defect}\approx 2G_0$ for $\varepsilon-\varepsilon_\text{F}$ about $-0.05$\,eV, as depicted in Figs.\,\ref{fig:conduct}(b) and (c), thus indicating that the edge-edge interaction is nearly suppressed around $0.05$\,eV below the Fermi level. Other (non-magnetic) defects may result in different energy intervals where the inter-edge coupling is minimized. On the other hand, in the limit of ${W}\rightarrow\infty$, perfect conducting channels with $G_\text{defect}=G_\text{prist}$ should be observed, independent of the scattering region length and the concentration of the defects. 

%%%%%%FIG6
 
\subsubsection{Localization length}

Given the random distribution of the scattering potential, here induced by the antistructure defects, at first glance our results of conductance can be interpreted in terms of Anderson localization. In this case, we can gain further insights by looking into the localization length $\xi$ as a function of the energy, where $\xi$ is defined as the length in which the conductance decays by an exponential factor \cite{RPPhyskramer1993, NANOTECHrocha2010}
\begin{equation}
G = G_0 e^{-L/\xi},
\end{equation} 
here, $G_0$ is an exponential prefactor, and $L$ is the length of scattering region. From the last equation, we can have
\begin{equation}
    \ln (G/G_0) = -\frac{L}{\xi}.
    \label{eq2}
\end{equation}

Within this model, $\xi$ is inversely proportional to the density of the scattering centers, $\rho$, namely $\xi\propto \rho^{-1}$ \cite{NANOTECHrocha2010}. Thus, it is expected that a given materials' NRs with the same defect concentration would present similar localization lengths. However, this is not what we found in our systems. Instead, $\xi$ presents a NR's width dependence due to ($i$) the inter-edge scattering process of the topological states and ($ii$) the distinct behavior of the electronic transport through the bulk and edge channels, where the latter (former) is (is not) topologically protected.

The behavior of $G$ vs $L$ is presented in Fig.~\ref{fig:conduct} where we estimated the localization lengths. Given the random distribution of defects for each $L$ an oscillation of $G/G_0$ for each energy is expected, that is, driven by the defect-defect. However across energies ranges wider than the defect-defect interaction a mean localization length is robust. In Fig.~\ref{fig:conduct}(f) we present average values of $\xi$ for energy ranges with $0.05$\,eV window. For energies outside the topological edge state range $\varepsilon-\varepsilon_{\rm F}<-0.1$ we see a localization length of $10^{-1}$\,{$\mu$m}, that is the localization of the system trivial bulk-like states. Within the topological gap $0.05>\varepsilon-\varepsilon_{\rm F}>-0.1$, the localization length increases in relation to those on the bulk-like energy range, that is, the backscattering forbidden topological transport is dominant. Additionally, within the topological edge states energy range there is an increase in the localization length with respect to the ribbon width $W$ reaching up to $10^{1}$\,{$\mu$m} for $W=99$\,{\AA} at the Fermi energy, $\varepsilon-\varepsilon_{\rm F} = 0$. Despite for narrow ribbons ($W=34$\,{\AA}) the non-magnetic impurity leads to edge-to-edge scattering; there exists energy ranges that can preserve long-range transport with conductance $2G_0$, for instance for $\varepsilon-\varepsilon_{\rm F} = 0$\,eV below the Fermi energy, reaching localization length up to $\sim 6$\,$\mu$m. Thus, regardless of intrinsic defects, the Pt$_2$HgSe$_3$ platform allows exploiting the topological states in narrow ribbons.

\section{Conclusions}

We investigated the electronic transport properties of 2D-Jacutingaite nanoribbons including defects and disorders within the systems. We analyzed the electronic conductance for different-width nanoribbons and different device lengths along the transport direction. Our ab initio results indicate a synergy between the confinement potential and manifestation of topological edge states that increase the energy range associated with the topological edge states in narrow ribbons. Stoichiometric defects in this system have a spatial length of $\sim 0.7$\,{nm}, while the edge state penetration depth, $\sim 0.9$\,{nm}, is much lower than in other 2D topological systems. By considering scattering lengths up to $109$\,nm, we found localization length that can surpass $\mu$m sizes for narrow nanoribbons ($<9$\,nm). This allows the exploitation of topological transport properties in narrow ribbons. In fact, by computing the transport localization length we show that a topological-driven transport can survive high defect densities $\sim 10^6$\,{cm$^{-1}$} for ribbons with width $W=34$\,{\AA}. Our analysis can contribute to the fundamental understanding and design integration of TIs within spintronic devices.

\section*{Conflicts of interest}
There are no conflicts to declare.

\begin{acknowledgments}
The authors acknowledge financial support from the Brazilian agencies FAPESP (grants 20/14067-3, 19/20857-0, and 17/02317-2), CNPq, INCT-Nanocarbono, INCT-Materials Informatics (grant 167651/2023-4), and Laborat\'{o}rio Nacional de Computa\c{c}\~{a}o Cient\'{i}fica for computer time (project ScafMat2 and emt2D).
\end{acknowledgments}

% \appendix

% \section{Appendixes}

\bibliography{bib}% Produces the bibliography via BibTeX.

\end{document}